\documentclass[a4paper,10pt]{article}

\usepackage{graphicx}

\usepackage{amssymb}

\begin{document}


\title{DC and AC Magnetization Study of Complex Ilmenite Oxides (Ni$_{1-x}$Co$_{x}$)TiO$_{3}$ ($0.05
\leq x \leq 0.8$)}

\author{Yukari Fujioka\footnote{Corresponding author: yukari.fujioka@aalto.fi} and Johannes Frantti\\Department of Applied Physics, \\Aalto University School of Science, P.O. Box 14100,
\\ 00076 Aalto, Finland}

\maketitle

\begin{abstract}

Ilmenite solid-solutions, (Ni$_{1-x}$Co$_{x}$)TiO$_{3}$ ($0.05 \leq
x \leq 0.8$), were synthesized at ambient atmosphere through
solid-state reaction and were studied by energy dispersive
spectroscopy of X-rays, X-ray diffraction, and DC and AC
magnetometry. Temperature dependent DC magnetic measurements
revealed two transitions. The first took place at around 27 K and
the second at 63 K. The low-temperature phase was antiferromagnetic.
The phase observed between 27 and 63 K is characteristic to the
solid-solution and is not found in either of the constituent
members. A fit of the data to the Curie-Weiss law gave magnetic
moment values which were significantly larger than the values based
on the quenched orbital momentum assumption.
Zero-field-cooled magnetization measurements with weak magnetic
field revealed unexpected negative magnetization at low
temperatures. Below 63 K the DC magnetization exhibited time
dependent behavior. Frequency and magnetic field dependent AC
magnetization is also addressed.
\end{abstract}

\section{Introduction}
\label{Introduction} NiTiO$_3$ (NTO) and CoTiO$_3$ (CTO) are both
antiferromagnetic below N\'eel temperatures 23 K\cite{Ishikawa} and
38 K\cite{Newnham}, respectively. They crystallize in ilmenite-type
structure with the space group $R\bar{3}$. Planes formed of double
sheets of triangular network of Ni (or Co) and Ti cations are
altered along the hexagonal $c$-axis and each cation is coordinated
by six oxygen atoms which form a distorted octahedron. The magnetic
moments of NTO and CTO order in the same way. Ni$^{2+}$ (Co$^{2+}$)
moments, parallel to the planes (the easy axis being perpendicular
to the $c$-axis), are ferromagnetically coupled within each plane
while the inter-plane coupling is antiferromagnetic
\cite{Newnham,Shirane}.

When ilmenite oxides form a solid-solution, complex and interesting
magnetic properties emerge. An example is
(Fe$_{y}$Co$_{1-y}$)TiO$_{3}$ (FCT), where the spin arrangement of
FeTiO$_3$ (FTO) is the same as that of NTO except for the spin
direction\cite{Shirane}. FCT is known to be a mixed Ising-\emph{XY}
antiferromagnet \cite{Ito82,Ito99}. The phase diagram of the
concentration ($x$) versus temperature ($T$) given in Ref.
\cite{Ito82} shows that there are three magnetically ordered phases
with respect to the composition: two antiferromagnetic phases with
different spin directions for Co- and Fe-rich areas and a 'mixed
ordering', or oblique antiferromagnetic, phase in the vicinity of
the Fe$_{0.5}$Co$_{0.5}$TiO$_3$ composition.

The motivation for the present study is to look a solid-solution
system closely related to FCT but in which the two constituents have
identical magnetic ordering. The study is dedicated on
polycrystalline (Ni$_{1-x}$Co$_{x}$)TiO$_{3}$ (NCT) solid solutions.
Our experiments reveal that, despite identical spin structures of
the constituents, magnetic properties of NCT behave in a complicated
manner.

\section{Material and methods}
Polycrystalline stoichiometric (Ni$_{1-x}$Co$_{x}$)TiO$_{3}$ ($0.05
\leq x \leq 0.8$) were prepared through the conventional solid-state
technique. Samples are referred to as NC$a$/$b$ indicating $a$\% of
Ni and $b$\% of Co. NiO (99 \%, Aldrich), CoO ($\geq$ 99.99 \%,
Aldrich) and TiO$_2$ (99 - 100.5 \%, Riedel-de Ha\"en) powders were
mixed in the desired molar ratio, then pressed into a pellet and
sintered in air at 1373 K. Phase quality was inspected by X-ray
diffraction using Cu K$\alpha$ radiation.

Composition analysis were carried out by energy dispersive
spectroscopy of X-rays (EDS) employing JEOL JSM-7500FA analytical
field-emission scanning electron microscope (SEM).

Magnetization measurements were performed by a superconducting
quantum interference device (SQUID) magnetometer (Quantum Design
XL7). DC magnetization in zero-field-cooled (ZFC) and field-cooled
(FC) runs were carried out over the temperature range down to 5 K
under various magnetic fields between 1 and 1000 Oe. Field-dependent
magnetization data as a function of applied field were collected at
various temperatures. AC magnetization measurements were performed
with the heating rate of 0.8 K/min with 1 Oe AC field
amplitude. Prior to measurements, sample was carefully centered with
respect to pick-up coils and the centering was regularly monitored.
Residual magnetic field was estimated using a paramagnetic material
prior to and after each measurement so that the real applied field
is given.

\section{Results and Discussion}
\subsection{Characterization}
\begin{figure}[f]
 \begin{center}
 \includegraphics[angle=0,width=8cm]{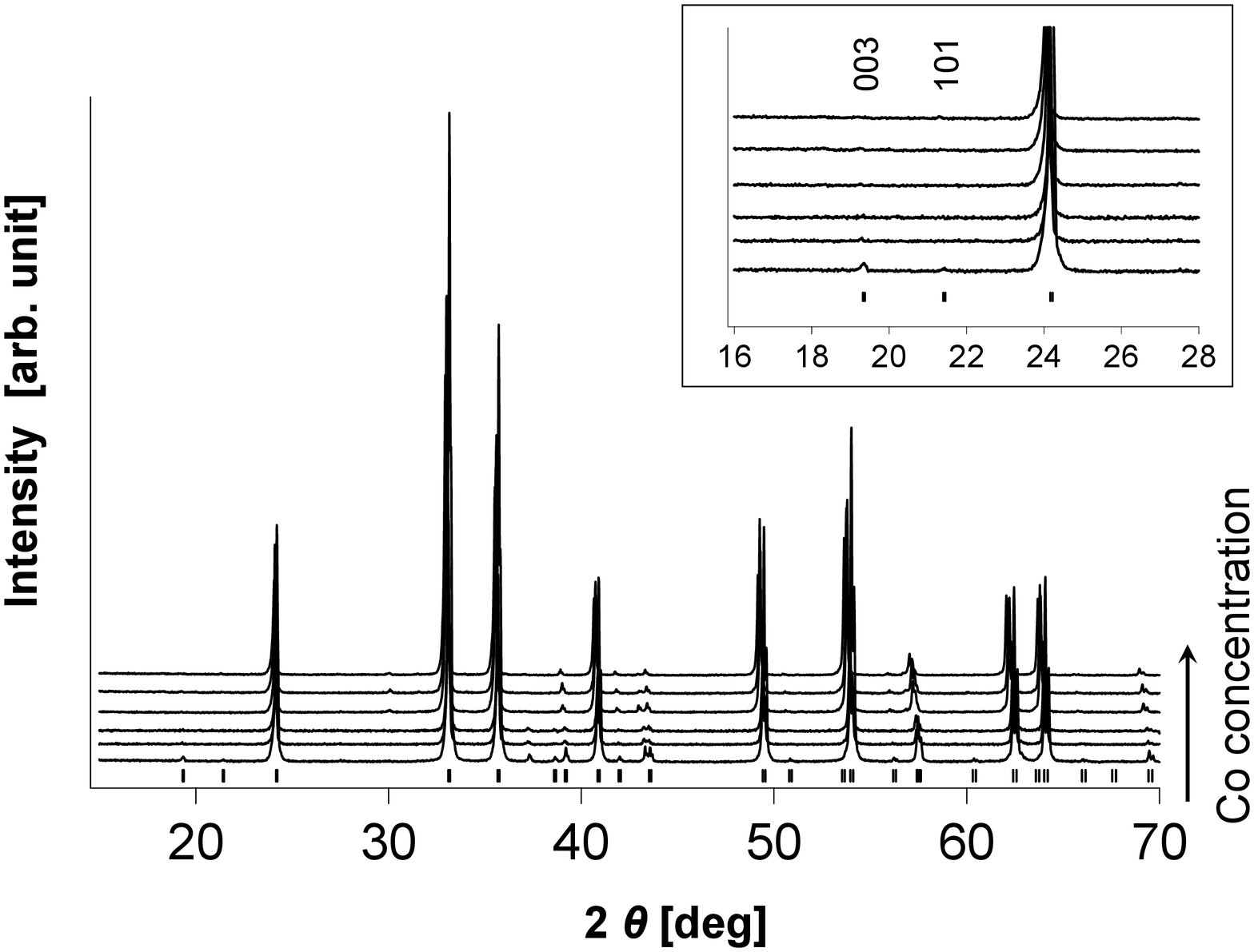}
 \end{center}
 \caption{Room-temperature XRD patterns collected on NCT powders.
 Samples are, from bottom to top, NC95/5, NC90/10, NC80/20, NC50/50,
 NC40/60, and NC20/80. Tick marks indicate reflections from the
 $R\bar{3}$ phase. Inset shows the 2$\theta$ region where the
 characteristic peaks of the $R\bar{3}$ phase are seen.} \label{XRD}
 \end{figure}

Because of the similarity in the crystal structure and chemical
character between NTO and CTO, one can expect solid solution with no
miscibility gap for whole composition range. In fact, all
solid-solutions in the tested composition range were obtained as
single phase. Figure \ref{XRD} shows X-ray diffraction patterns of
selected NCT powders. The NCT sample with lowest Co content was
light green in color and became dark with increasing Co content.
$h0l$ ($l$ is odd) reflections, such as 003 and 101, indicative to
the ilmenite phase $R\bar{3}$. Disappearance of these reflections
implies a higher symmetry $R\bar{3}c$. For NTO and CTO, however, the
$h0l$ peaks are usually rather weak and the degree of ordering could
not be determined from the XRD data alone. The room-temperature
X-ray diffraction also show that lattice parameters were increasing
with increasing Co-content, suggesting that Co$^{2+}$ and Ni$^{2+}$
were in a high-spin state (their ionic radii are $0.745$ \AA{} and
$0.690$ \AA{}, respectively\cite{Rohrer}, in contrast, the low-spin
ionic radii of Co$^{2+}$ is $0.65$ \AA{}\cite{Rohrer}). However, as
the antiferromagnetic ordering temperatures of CoTiO$_3$ and
NiTiO$_3$ demonstrate, the consideration of bond-lengths alone is
insufficient for understanding the composition dependent magnetic
transition temperatures.

Homogeneity of the compounds were also inspected by SEM-EDS. No
spatial composition variation was evidenced. Compositions determined
by SEM-EDS, listed in Table \ref{Table1} show that the starting
molar ratio was preserved well.

\subsection{Magnetic properties}
\begin{figure}[f]
 \begin{center}
 \includegraphics[angle=0,width=8.25cm]{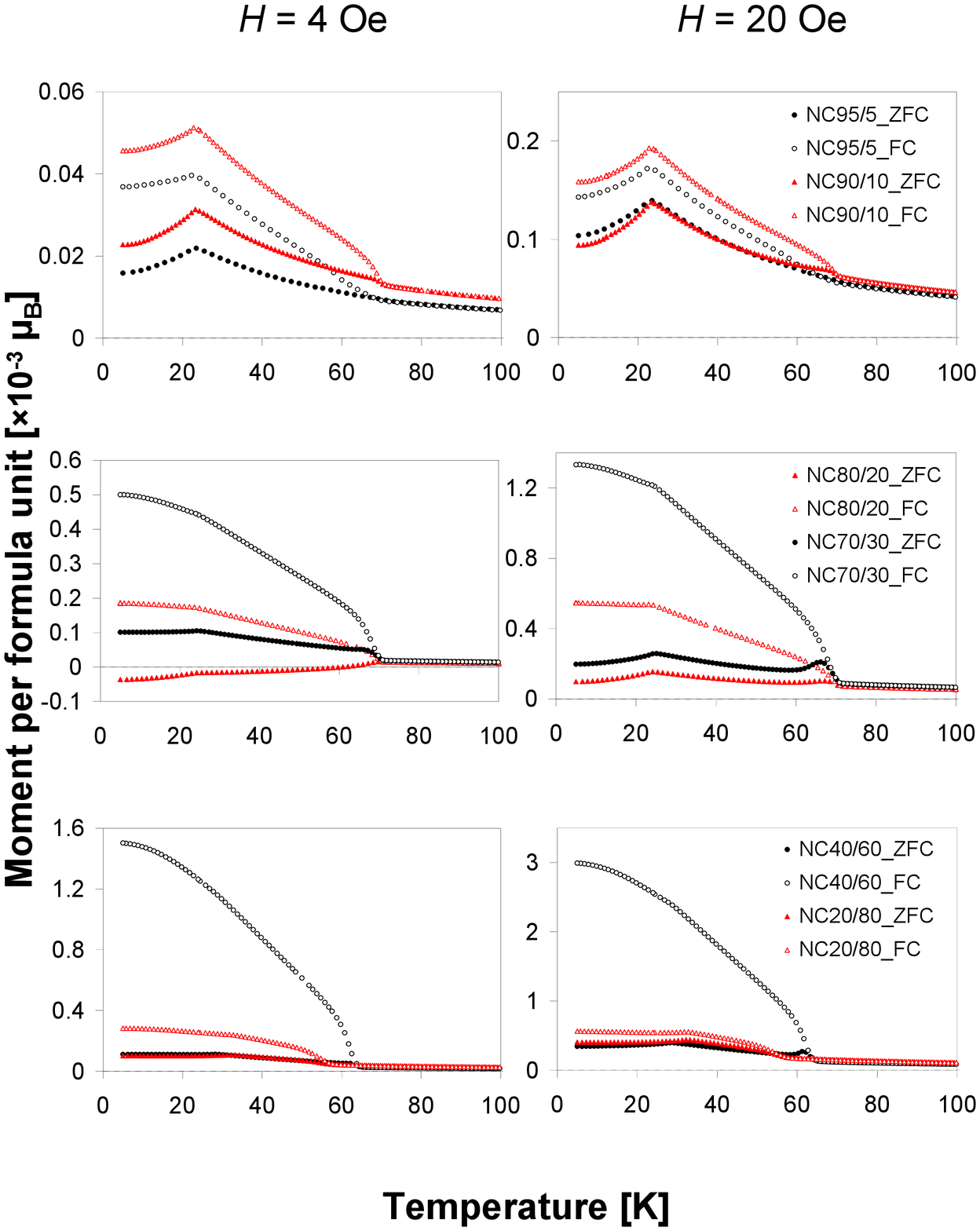}
 \end{center}
 \caption{Temperature and weak field dependent magnetization. Co
 concentration increases from top to bottom. Panel pairs at the same
 row show data from the same sample. Filled and open marks indicate
 ZFC and FC data, respectively.}\label{MTL}
  \end{figure}

\begin{figure}[f]
 \begin{center}
 \includegraphics[angle=0,width=8.25cm]{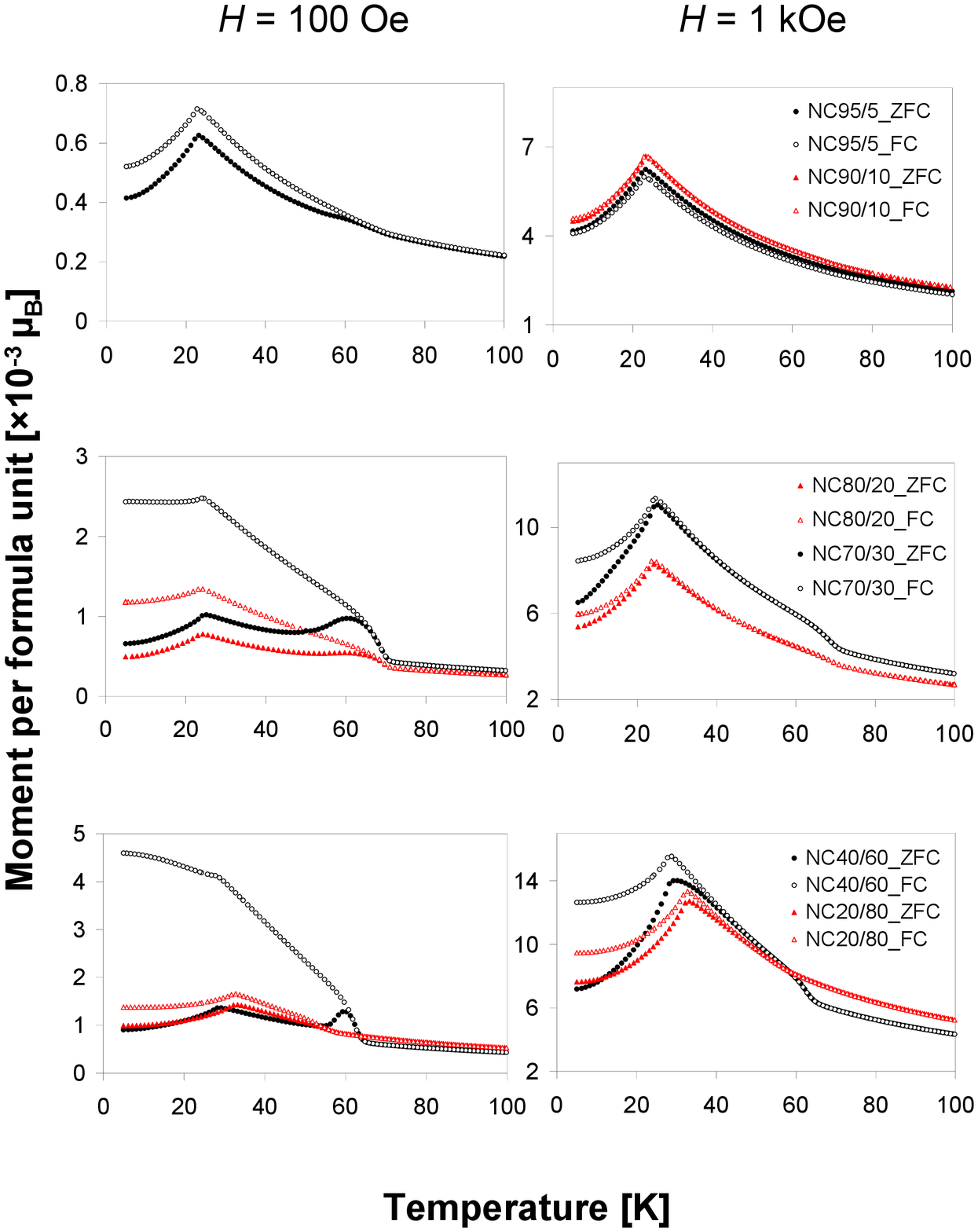}
 \end{center}
 \caption{Temperature and field-dependent magnetization. Co
 concentration increases from top to bottom. Panel pairs at the same
 row show data from the same sample. Filled and open marks indicate
 ZFC and FC data, respectively.} \label{MTH}
 \end{figure}

\paragraph{Overview of DC magnetization results.}
Figures \ref{MTL} and \ref{MTH} show the temperature dependence of
DC magnetization [$M(T)$] at various applied fields. Overall the
magnetic properties of NCT samples were quite similar: there are two
magnetic transitions at around 23 and 67 K. The higher temperature
transition is only characteristic to the solid-solutions. Above this
temperature, NCTs were paramagnetic: the inverse susceptibility 
was linearly decreasing with decreasing temperature down to the transition temperature.

In the paramagnetic phase, susceptibility obeys the Curie-Weiss law.
Correspondingly, effective magnetic moments $\mu_{e}$ were estimated
from the paramagnetic region data, as shown in Table \ref{Table1}.
For comparison, the magnetic moment values were estimated from the
equation based on the assumption of quenched orbital moments:
$\mu_{e,q} = 2\{(1-x)\sqrt{S_{Ni^{2+}}(S_{Ni^{2+}}+1)} +
x\sqrt{S_{Co^{2+}}(S_{Co^{2+}}+1)}\}$, where $S_{Ni^{2+}}$ and
$S_{Co^{2+}}$ are total electron spin moment values of Ni and Co
ions in their high-spin states, respectively. The difference between
the magnetic moments estimated from the susceptibility data and the
quenched orbital moment values became larger with increasing Co
content, suggesting that the orbital moments of Co-ions were not entirely
quenched. Interestingly, the magnetic moment values of NiTiO$_3$
\cite{Ishikawa} and notably of CoTiO$_3$ \cite{Newnham} were larger than the
values expected for quenched orbital moments. For comparison, also
values based on the Hund's rules are given in Table \ref{Table1}. 
Also the Co$^{2+}$ ions in LaMn$_{0.5}$Co$_{0.5}$O$_3$ were found 
to possess a large orbital moment, which was predicted to result in a 
nontrivial temperature dependence of the magnetic susceptibility \cite{Burnus}.

\begin{figure}[f]
 \begin{center}
 \includegraphics[angle=0,width=8.25cm]{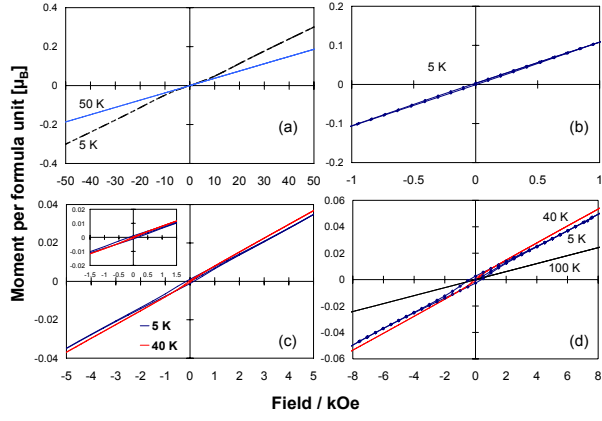}
 \end{center}
 \caption{Field-dependent magnetization at constant temperature of
 the samples with $x \leq 0.3$. (a)NC95/5, (b)NC90/10, (c)NC80/20,
 (d)NC70/30.} \label{MHL}
 \end{figure}

\begin{figure}[f]
 \begin{center}
 \includegraphics[angle=0,width=8.25cm]{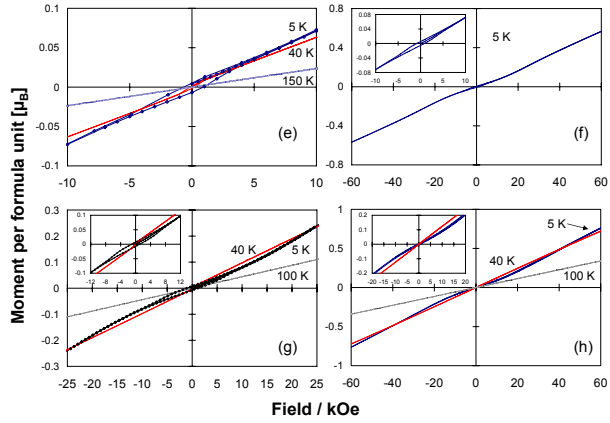}
 \end{center}
 \caption{Field-dependent magnetization at constant temperature of
 the samples with $0.4 \leq x \leq 0.8$. (e)NC60/40, (f)NC50/50,
 (g)NC40/60, (h)NC20/80.} \label{MHH}
 \end{figure}

The low temperature transition was close to the $T_N$ of NTO. With
increasing field, the cusp at around 27 K became sharper. The cusp
is characteristic signal to a transition to an antiferromagnetic
phase. The peak temperature of the cusp is referred to as
$T_{\textrm{AFM}}$. Antiferromagnetic order is consistent with the
negative values of asymptotic Curie point $\theta_{a}$/K and the
results of field dependent magnetization [$M(H)$], as given in Figs.
\ref{MHL} and \ref{MHH}. Magnetization was almost linearly dependent
of the applied field, except for the small hysteresis loop which
diminished significantly at 40 K and vanished above the high
temperature transition. The opening was largest at around $x=0.50$
composition, and decreased for small and large values of $x$. With
increasing Co content up to 0.8, $T_{\textrm{AFM}}$ almost linearly
increased to 32 K. The linear increase of $T_{\textrm{AFM}}$ with
increasing $x$ parallels with the linear increase of magnetic moment
$\mu_{e}$, see Table \ref{Table1}. Despite the fact that the solid
solution has larger number of magnetic interactions (Ni-Ni, Ni-Co
and Co-Co interactions in simplest terms) than the parent compounds,
it seems that the exchange integrals $\mathit{J}$ of different
magnetic atom pairs are roughly the same, in contrast to
(Fe$_{y}$Co$_{1-y}$)TiO$_{3}$ or other spin glass compounds. For a
typical spin glass material, glass transition temperature declines
when the composition approaches the intermediate
region\cite{Landau}. This behavior is explained by random
distribution of $\mathit{J}$ over magnetic ion pairs.

We estimated the temperature at which the inverse DC susceptibility
versus temperature curve starts to depart from the extrapolation of
Curie-Weiss relation, the onset of a transition. Hereafter this
temperature is referred to as irreversible point $T_{irr}$.
$T_{irr}$ exhibited composition dependence: increasing Co content
pushed the transition temperature down. The decrease was not linear
as a function of Co content but was faster at highest Co content so
that $T_{irr}$ decreased to 57 K for NC20/80 sample. $T_{irr}$ was
sensitive to an applied field: with increasing applied field 
$T_{irr}$ was shifted to lower temperature.

\begin{figure}[f]
 \begin{center}
 \includegraphics[angle=0,width=8cm]{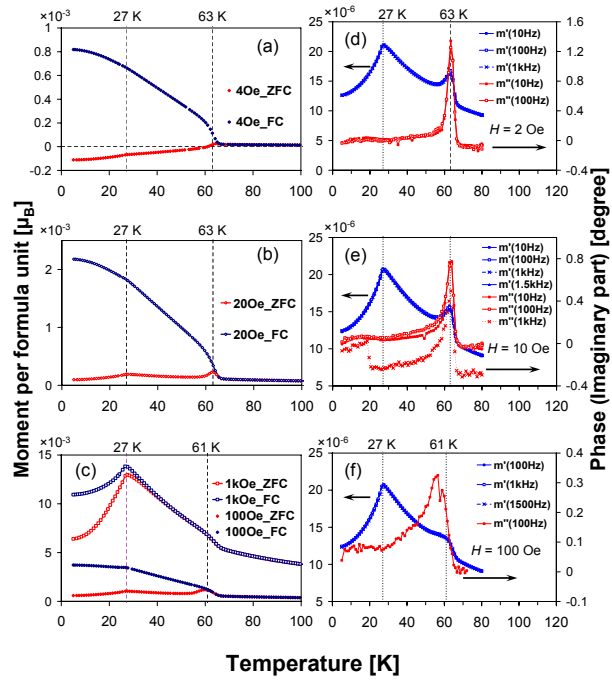}
 \end{center}
 \caption{Left column shows temperature dependent DC magnetization of
 the NC50/50 sample. Right column shows field and frequency dependent
 AC magnetization. For AC imaginary part, phase difference in degrees is given in the second vertical axis.
Peak temperatures of AC magnetization curves are
 indicated by dashed lines. It is worth to note that the data
 measured with different frequencies overlap, in contrast to spin
 glass systems.} \label{NC55}
 \end{figure}

\begin{figure}[f]
 \begin{center}
 \includegraphics[angle=0,width=8cm]{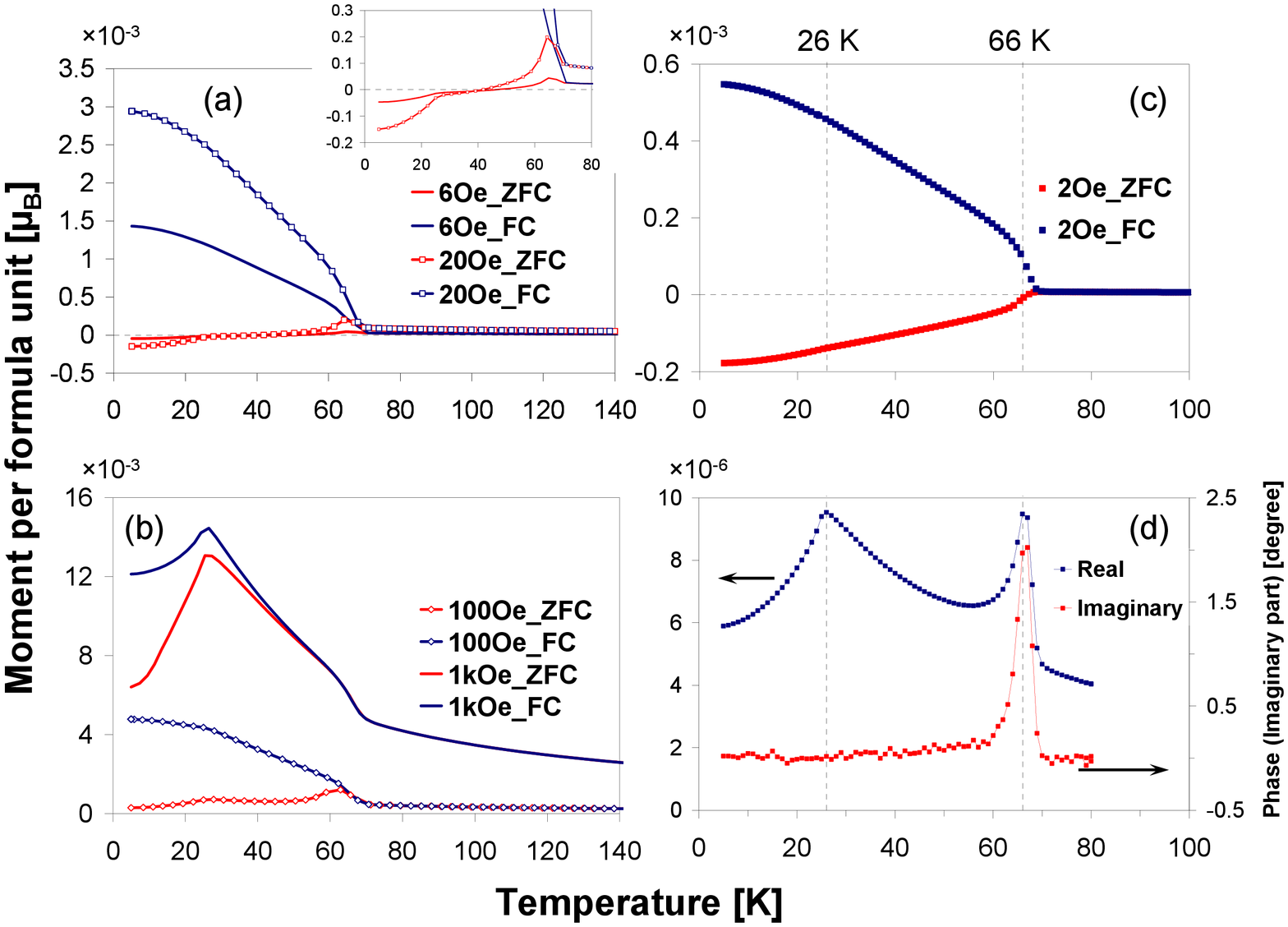}
 \end{center}
 \caption{Temperature dependent DC and AC magnetization of the
 NC60/40 sample. In panel (a), closer look around zero magnetization
 is given in inset. AC data (d) was measured using a DC field 1.5 Oe,
 AC-field amplitude 1 Oe and frequency 10 Hz.} \label{NC64}
 \end{figure}

\paragraph{DC and AC magnetization measurements of NC50/50 and NC60/40.}
To look closer to the nature of the transitions, frequency
dependence of AC magnetization with various applied fields were
measured for NC50/50 and NC60/40. In both cases, real and imaginary
AC magnetization showed sharp peaks at the point corresponding to
the high temperature transitions, see Figs. \ref{NC55} and
\ref{NC64}. $T_{\textrm{AFM}}$ also corresponds to the low
temperature cusp of real part of AC magnetization curves. Next we
focus on the result of NC50/50, although the discussion is valid to
all NCT solid-solutions.

With increasing field, the difference between ZFC and FC curves
[panels (a), (b), and (c) in Fig. \ref{NC55}] became smaller and the
higher temperature transition slightly shifted to lower temperature,
as was commonly observed in all other NCTs. The higher temperature
transition at which both the real and imaginary parts of AC
magnetization curves show a peak is labeled as $T_p$. We note that
$T_p$ is slightly lower than $T_{irr}$, though they indicate the
same magnetic transition. Except for the peak at $T_p$, the
imaginary part of magnetization was close to zero. A sharp peak in
imaginary part generally indicates that magnetic domain structure with
switchable magnetization is formed.
On the contrary, clear yet small hysteresis loop was observed only at low temperature: the net magnetization was largest at lowest
temperature. Thus, the dissipation peak at $T_p$ and the ferromagnetic hysteresis 
seem to be unrelated phenomena suggesting the different nature of the two phases.  The small hysteresis loop can be explained by canting the
antiferromagnetically ordered spins so that there is a small net
magnetization. 
Both DC and AC magnetization results of the low-temperature phase are consistent with the canted antiferromagnetic order.
The difference in magnetic ordering was also reflected by the different shape of the 
magnetization versus applied field: whereas the lowest temperature curve had an s-shape, 
above $T_{AFM}$ the curve was pratically a straight line.


The dissipation peak at $T_p$ suggests a formation of
domains which are sufficiently large to show no frequency dependent
response. Consistently with the DC magnetization results, the $T_p$, 
determined through the AC-measurements, 
was slightly shifted with increasing applied field. 
This suggests
that the origin of the phenomena is on the collective spin
ordering on the layers perpendicular to the hexagonal $c$-axis
rather than magnetic nanodomains. Specifically the high $T_{p}$ 
and the sensitivity to an applied field further implies that
the magnetic phase transitions can be related to the in-plane spin
order.

\begin{figure}[f]
 \begin{center}
 \includegraphics[angle=0,width=5cm]{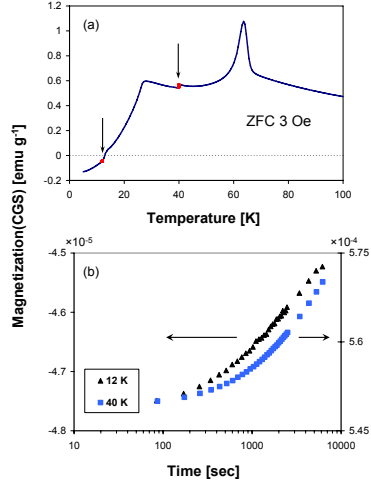}
 \end{center}
 \caption{(a) DC magnetization collected continuously on
 zero-field-cool cycle with the magnetic field of 3 Oe from the
 NC50/50 sample. At 12 and 40 K (indicated by arrows), temperature
 was kept constant for 100 min. (b) Time dependent magnetization at
 12 and 40 K.} \label{TD}
 \end{figure}

The magnetization of the NC50/50 samples in ZFC curve measured with
weak field was negative up to $T_p$, see panel (a) in Fig.
\ref{NC55}. Negative magnetization was also observed in the NC80/20
and NC60/40. The temperature at which the sign of magnetization
changes varied by the heating rate. To verify the time dependence of
the magnetization, data were collected keeping the temperature
constant at 12 and 40 K for 100 min in ZFC run with weak external
field, as shown in Fig. \ref{TD}. Magnetization curve in Fig.
\ref{TD}(a) crosses zero at much lower temperature comparing to Fig.
\ref{NC55}(a), still the peak temperatures matched to $T_{AFM}$ and
$T_p$. With increasing time, magnetization at 12 and 40 K gradually
increased though the change at 40 K was larger than 12 K. Similar
time-dependent magnetization can be seen in a spin glass. However,
AC magnetization data (the right column of Fig. \ref{NC55})
contradict the characteristic feature of spin glass\cite{Landau}
which displays a clear frequency dependence when a small field is
applied: neither real or imaginary part of AC magnetization curves
exhibited frequency dependence. The construction of a detailed picture
of the magnetic and possible cation ordering requires neutron and
synchrotron diffraction studies.

\begin{table}[f]
\caption{Compositions, N\'eel point $T_{\textrm{AFM}}$/K,
irreversible point $T_{irr}$/K, effective moment $\mu_{e}$ in Bohr
magneton unit $\mu_{B}$ and asymptotic Curie point $\theta_{a}$/K
estimated from Curie-Weiss law and effective moment $\mu_{e,q}$ in
Bohr magneton unit assuming orbital quenching of
(Ni$_{1-x}$Co$_{x}$)TiO$_{3}$ ($0.05 \leq x \leq 0.8$). Also moments
$\mu_{e,J} = g_J\sqrt{J(J+1)}$, where $J$ is the total orbital
angular moment and $g_J$ is the Land\'e g-value (based on Hund's
rules expectations) are given. Compositions indicate cation ratio
(oxygen content was not determined).} \label{Table1}
\begin{center}
\begin{tabular}{l l l l l l l l}
\hline
Compound    &  Composition                        & $T_{\textrm{AFM}}$ & $T_{irr}$ &$\mu_{e}$ & $\theta_{a}$ & $\mu_{e,q}$ & $\mu_{e,J}$\\
NiTiO$_{3}$\cite{Ishikawa} &                             & 23       & --        &  3.24      & -55   & 2.87     & 5.59 \\
NC95/5      & (Ni$_{0.95}$Co$_{0.05}$)Ti$_{1.00}$O$_{3}$ & 23       & 71        &  3.00      & -17   & 2.88     & 5.64 \\
NC90/10     & (Ni$_{0.90}$Co$_{0.11}$)Ti$_{0.99}$O$_{3}$ & 23       & 70        &  3.58      & -19   & 2.97     & 5.69 \\
NC80/20     & (Ni$_{0.77}$Co$_{0.23}$)Ti$_{1.00}$O$_{3}$ & 24       & 69        &  3.80      & -19   & 3.07     & 5.80 \\
NC70/30     & (Ni$_{0.68}$Co$_{0.33}$)Ti$_{0.99}$O$_{3}$ & 24       & 71        &  4.32      & -18   & 3.20     & 5.90 \\
NC60/40     & (Ni$_{0.62}$Co$_{0.40}$)Ti$_{0.97}$O$_{3}$ & 25       & 69        &  4.03      & -16   & 3.30     & 6.01 \\
NC50/50     & (Ni$_{0.53}$Co$_{0.47}$)Ti$_{0.99}$O$_{3}$ & 26       & 66        &  4.21      & -15   & 3.32     & 6.11 \\
NC40/60     & (Ni$_{0.42}$Co$_{0.60}$)Ti$_{0.99}$O$_{3}$ & 28       & 66        &  4.73      & -14   & 3.51     & 6.22 \\
NC20/80     & (Ni$_{0.19}$Co$_{0.81}$)Ti$_{0.99}$O$_{3}$ & 32       & 57        &  5.77      & -11   & 3.67     & 6.42 \\
CoTiO$_{3}$\cite{Newnham} &                              & 38       & --        &  5.3       & -15   & 3.87     & 6.63 \\
\hline
\end{tabular}
\end{center}
\end{table}

\section{Conclusions}
In conclusion, solid-solution of NiTiO$_3$ and CoTiO$_3$,
(Ni$_{1-x}$Co$_x$)TiO$_3$ ($0.05 \leq x \leq 0.80$), were
successfully prepared through the solid-state reaction at ambient
atmosphere. DC and AC magnetization revealed that below the room
temperature NCT solid solution has two magnetic transitions: to an
antiferromagnetic phase, which took place $T_{AFM}$ = 27 K and
another transition which occurred around $T_{p}$ = 63 K, being
significantly higher than $T_{AFM}$. Small net magnetization was 
observed at the lowest temperatures, which was probably due to 
the canted spins. Negative magnetization was
observed at weak fields for samples with $x = $ 0.2, 0.4 and 0.5.
Unlike in most $3d$ transition metal oxides, the orbital moments of Ni 
and Co in NCT were not entirely quenched. Though the time
dependence of magnetization was observed, there was no 
evidence for a spin glass transition. In summary, we have shown that
peculiar magnetic properties emerge in NCT system.

\section{Acknowledgements}
We thank A. Savin for his help in operating the magnetometer.





\bibliographystyle{elsarticle-num}



\end{document}